\documentclass[11pt]{article}
\usepackage{amsmath}
\usepackage{amssymb}
\usepackage{cite}

\begin{document}

\newcommand{\pst}{\hspace*{1.5em}}
\setcounter{footnote}{0} \setcounter{equation}{0}
\setcounter{figure}{0} \setcounter{table}{0} \setcounter{section}{0}

\newcommand{\be}{\begin{equation}}
\newcommand{\ee}{\end{equation}}
\newcommand{\bm}{\boldmath}
\newcommand{\ds}{\displaystyle}
\newcommand{\bea}{\begin{eqnarray}}
\newcommand{\eea}{\end{eqnarray}}
\newcommand{\ba}{\begin{array}}
\newcommand{\ea}{\end{array}}
\newcommand{\arcsinh}{\mathop{\rm arcsinh}\nolimits}
\newcommand{\arctanh}{\mathop{\rm arctanh}\nolimits}
\renewcommand{\thefootnote}{\fnsymbol{footnote}}
\newcommand{\bc}{\begin{center}}
\newcommand{\ec}{\end{center}}
\newcommand{\degree}{$^\circ$}
\newcommand{\blackcircle}{{\begin{picture}(6,12)\put(3,3){\circle*{6}}\end{picture}}}

\setcounter{footnote}{0} \setcounter{equation}{0}
\setcounter{figure}{0} \setcounter{table}{0} \setcounter{section}{0}

\thispagestyle{plain}
\label{sh}

\addtocounter{footnote}{1}
\noindent\textbf{\Large \tabcolsep=0mm
\begin{tabular}{c}
\rule{175mm}{0mm}\\[-6mm]
HIDDEN QUANTUM CORRELATIONS\\[-1mm]
IN SINGLE QUDIT SYSTEMS\footnote{}
\end{tabular}}

\footnotetext{Based on the talk by M.A. Man'ko at
the 22nd Central European Workshop on Quantum Optics (6--10 July
2015, Warsaw, Poland).}

\bigskip

\begin{center} {\bf
Margarita A. Man'ko$^{1\,\ast}$ and Vladimir I. Man'ko$^{1,2}$}

\medskip

{\it
$^1$Lebedev Physical Institute, Russian Academy of Sciences\\
Leninskii Prospect 53, Moscow 119991, Russia
\smallskip

$^2$Moscow Institute of Physics and Technology (State University)\\
Institutski\'{\i} per. 9, Dolgoprudny\'{\i}, Moscow Region 141700, Russia }
\smallskip

$^*$Corresponding author e-mail:~~~mmanko\,@\,sci.lebedev.ru\\
\end{center}

\begin{abstract}\noindent
We introduce the notion of hidden quantum correlations. We present
the mean values of observables depending on one classical random
variable described by the probability distribution in the form of
correlation functions of two (three, etc.) random variables
described by the corresponding joint pro\-bability distributions. We
develop analogous constructions for the density matrices of quantum
states and quantum observables. We consider examples of
four-dimensional Hilbert space corresponding to the ``quantum
roulette'' and ``quantum compass.''

\end{abstract}

\medskip


\section{Introduction}
\pst Quantum correlation phenomena, like the entanglement~\cite{Schroedinger}
present in composite systems, for example, in the system of several
qubits, are known to play an important role in developing new
quantum technologies, including quantum computing~\cite{Chuang}.
Strong quantum correlations in two-qubit systems responsible for the
violation of Bell inequalities~\cite{Bell,CHSH} were checked
experimentally~\cite{Aspect,Grange}. In \cite{Shumovsky}, it was
suggested to extend the notion of entanglement in order to relate
this phenomenon to correlation properties of single qudits.

The new entropic inequalities reflecting the presence of
correlations and analogous to the subadditivity and strong
subadditivity conditions known for bipartite and tripartite
systems~\cite{Lieb,Ruskai,Petz} were found for noncomposite systems
like single qudits or multilevel
atoms~\cite{11,12,13,13a,13b,Entropy,13d}. Examples of qudits,
including $j=3/2$, were considered in this context in
\cite{14,15,16,Olga,Markovich}, and the results obtained show that the
correlations in composite systems and the correlations in
noncomposite systems can formally be considered as identical, using
a common mathematical framework.

The aim of this work is to develop the approach for describing both
classical and quantum correlations in composite and noncomposite
systems, using the same scheme based on the application of
invertible maps of integer numbers $s$ onto pairs (triples, etc.) of
the integers $(j,k)$ employed in \cite{Entropy}. Employing these
maps, we demonstrate that a single variable and its statistical
properties, such as mean values, can be considered as the properties
of several random variables described by the corresponding joint
probability distributions and given in terms of the correlation
functions calculated for these several random variables. We show
this property for both classical and quantum systems.

This paper is organized as follows.

In Sec.~2, we discuss the means and correlations in classical
systems. In Sec.~3, we present examples of four- and
eight-dimensional probability distributions. In Sec.~4, we study
quantum states, and in Sec.~5 we consider in detail the case of
$N=4$ along with entropic inequalities. We give our conclusions in
Sec.~6.

\section{Means as Correlation Functions}
\pst
Our aim now is to consider correlations in a single qudit as
correlations in artificial multiqudit systems. We start from
classical states.

Following~\cite{11,Entropy} we consider a set of nonnegative numbers
$A_1,A_2,\ldots,A_N$ which, in turn, provides a set of other
nonnegative numbers $0\leq p_s=\dfrac{A_s}{\sum_{j=1}^NA_j}\leq 1$
satisfying the normalization condition $\sum_{s=1}^Np_s=1.$ The
numbers $p_s$ can be interpreted as the probability distributions of
one random variable.

Let us measure the observable $F(s)$. For each value of the integer
$s=1,2,\ldots,N$, one obtains the result of the measurement $F(s)$.
Repeating the measurement $L$ times, where $L$ is a large enough
integer, one obtains such statistical characteristic as the mean
value of the measured observable
\begin{equation}\label{1}
\langle F\rangle=\sum_{s=1}^Np_sF(s).
\end{equation}
We consider $s$ as {\it a random variable}, $F(s)$ as {\it an
observable}, and $p_s$ as the probability distribution of one random
variable, which we call {\it the state}. The other statistical
characteristics described by the highest moments like, for example,
variances
\begin{equation}\label{2}
\langle F^2\rangle -\langle F\rangle^2=\sum_{s=1}^Np_sF^2(s)-
\Big(\sum_{s=1}^Np_sF(s)\Big)^2,
\end{equation}
can also be obtained.

Formally, one has two functions $F(s)$ and $p_s$ defined on the set
of integers $s=1,2,\ldots,N$.

We call the function $p_s$ {\it the state} and the function $F(s)$
{\it the observable} due to the following reason.

If one has the continuous variable $x$ $\big($a position of the
particle with the Gaussian distribution $P(x)\big)$, we call $x$ the
random variable and the distribution $P(x)$ the state of the system.
We extend this terminology to a discrete variable $s$ and the
distribution $p_s$.

The mean value $\langle F\rangle$ and variance $\langle F^2\rangle
-\langle F\rangle^2$ are the functionals given by Eqs.~(\ref{1}) and
(\ref{2}). In principle, one can formally define the functions
$F(s)$ and $p_s$, as well as the functionals (\ref{1}) and
(\ref{2}), without the probabilistic interpretation of these
objects.

On the other hand, there exist functionals determined not by both
functions $F(s)$ and $p_s$ but by only one function $p_s$. For
example, Shannon entropy~\cite{Shannon} associated with the
probability distribution $p_s$ is given by the expression
\begin{equation}\label{3}
H=-\sum_{s}p_s\ln p_s;
\end{equation}
the entropy being the functional of the state.

Meanwhile, the entropy $H$ is the functional which can also be
considered formally without a pro\-babilistic interpretation of the
numbers $p_s$. We point out the possibility to treat the functionals
$\langle F\rangle$, $\langle F^2\rangle -\langle F\rangle^2$, and
$H$ as objects that can be considered without their probabilistic
interpretation because the numerical properties of these and other
analogous functionals, e.g., all highest moments
\begin{equation}\label{4}
\langle F^k\rangle=\sum_{s=1}^Np_sF^k(s),
\end{equation}
like equalities and inequalities for these objects, exist
independently of their relation to probabilities.

One can repeat the above consideration for nonnegative numbers
$p_{jk}$. This simple observation can be used for obtaining some new
equalities and inequalities for functionals (entropies,
correlations, means, variances, and covariances) associated with
tables of nonnegative numbers $0\leq p_{jk}\leq 1$,
$j=1,2,\ldots,n$, $k=1,2,\ldots,m$, and $N=nm$, since the table can
be considered as a joint probability distribution for two random
variables.

Within the framework of the interpretation of the table $p_{jk}$ as a joint
probability distribution, the characteristics like entropy, mutual
information, etc. naturally appear. These characteristics are known to satisfy
the entropic inequalities for bipartite classical systems; see \cite{Holevo}.

On the other hand, the numerical expressions of these inequalities
are valid independently of the probabilistic interpretation of the
numbers in the table $p_{jk}$. We employ this fact for obtaining new
inequalities for {\it the state} (the probability distribution
$p_s$) associated with one random variable and the function $F(s)$
$\big(${\it observable} $F(s)\big)$. The key tool to achieve this
result is introducing the map of integers, namely, for
$s=1,2,\ldots,N$ we construct the invertible map
$$1\leftrightarrow 1,1;~~2\leftrightarrow 2,1;~~\ldots;~~n\leftrightarrow
n,1;~~n+1\leftrightarrow 1,2;~~n+2\leftrightarrow 2,2;~~\ldots;
N-1\leftrightarrow n-1,m;~~N\leftrightarrow n,m.$$ This map could be
described as a procedure for introducing the function $s(j,k)$. Such
a function defined in the domain of integers $j=1,2,\ldots,n$ and
$k=1,2,\ldots, m$ provides for each pair of the integers $j,k$ the
value of the function equal to the integer $s$. The function is
constructed using the invertibility condition; this means that for
each value of the integer $s$ one has only one pair of integers
$j,k$ corresponding to this value. Such construction was used in
\cite{Entropy} to derive new entropic inequalities for qudit states.
Here, we extend this construction to study the properties of
observables associated with functions $F(s)$.

In fact, these observables associated with one random variable can
be treated as observables connected with two random variables. To
demonstrate this fact, we define the function $\Phi(j,k)\equiv
F\big(s(j,k)\big)$. One can choose this function in the product form
\begin{equation}\label{5}
\Phi(j,k)=\phi(j)\chi(k).
\end{equation}
The form of ${\it {observable}}$ $F\big(s(j,k)\big)$ provides the
possibility to interpret the observable as the existence of two
${\it {observables}}$ $\phi(j)$ and $\chi(k)$ associated with two
random variables $j$ and $k$. Also the probability distribution
$p_s$ can be chosen in the product form $p_{jk}=\Pi_j{\cal P}_k$;
this representation can be chosen with high ambiguity. In view of
this representation, one can rewrite formula~(\ref{1}) for the mean
value $\langle F\rangle$ as follows:
\begin{equation}\label{6}
\langle
F\rangle=\sum_{s=1}^Np_sF(s)=\sum_{j=1}^n\sum_{k=1}^m\Phi(j,k)p_{jk}.
\end{equation}
Also one can introduce marginal probability distributions:
\begin{eqnarray}
\Pi_j=\sum_{k=1}^mp_{jk}\equiv\sum_{k=1}^mp_{s(j,k)},\qquad
j=1,2,\ldots,n,\label{7}\\
{\cal P}_k=\sum_{j=1}^np_{jk}\equiv\sum_{j=1}^np_{s(j,k)},\qquad
k=1,2,\ldots,m.\label{8}
\end{eqnarray}
If the numbers $p_s$ determining the joint probability distribution
are such that $p_{jk}=\Pi_j{\cal P}_k$, where
$\sum_{j=1}^n\Pi_j=\sum_{k=1}^m{\cal P}_k=1$, one has for the
marginal distributions~(\ref{7}) and (\ref{8}) the case of the
absence of correlations between the observables associated with the
function $\Phi(j,k)$ given by (\ref{5}).

If the function $F\big(s(j,k)\big)$ has the product form analogous
to (\ref{5}), the mean value of this function $\langle F\rangle$
given by (\ref{6}) and written as
\begin{equation}\label{9}
\langle F\rangle=\sum_{j=1}^n\sum_{k=1}^m\phi(j)\chi(k)p_{jk}
\end{equation}
can be interpreted as the correlation function, i.e.,
\begin{equation}\label{10}
\langle F\rangle=\langle \phi(j)\chi(k)\rangle.
\end{equation}
Thus, for one random variable $F(s)$ we obtain the formula for its
mean value in the form of correlation function associated with two
observables depending on random variables $\phi(j)$ and $\chi(k)$,
using the averaging procedure determined by the joint probability
distribution $p_{jk}$.

In the case of integer $N=n_1n_2n_3$, where the factors in the
product are integers, one can use the invertible map of the integers
$s$ onto the triples of integers $j,k,\ell$, where
$j=1,2,\ldots,n_1$, $k=1,2,\ldots,n_2$, and $\ell=1,2,\ldots,n_3$.
This means that we construct the function of three variables
$s(j,k,\ell)$ such that for each three integers we have only one
integer $s$, and for each integer $s$ we have only one triple of
integers $j,k,\ell$.

In view of this invertible map, the probability distribution $p_s$
used to describe statistical properties of observable $F(s)$
depending on one random variable $F(s)$ may be interpreted as the
joint probability distribution $p_{s(j,k,\ell)}\equiv p_{jk\ell}$ of
three random variables. For this, we define the function
$T(j,k,\ell)\equiv F\big(s(j,k,\ell)\big)$, which can be chosen in
the product form
\begin{equation}\label{11}
T(j,k,\ell)=a(j)b(k)c(\ell).
\end{equation}
Then one can write the equality
\begin{equation}\label{12}
\langle
F\rangle=\sum_{s=1}^Np_sF(s)=\sum_{j=1}^{n_1}\sum_{k=1}^{n_2}\sum_{\ell=1}^{n_3}
T(j,k,\ell)p_{s(j,k,\ell)}
\end{equation}
or
\begin{equation}\label{13}
\langle
F\rangle=\sum_{j=1}^{n_1}\sum_{k=1}^{n_2}\sum_{\ell=1}^{n_3}
a(j)b(k)c(\ell)p_{s(j,k,\ell)},
\end{equation}
which means
\begin{equation}\label{14}
\langle
F\rangle=\langle a(j)b(k)c(\ell)\rangle.
\end{equation}
Thus, we presented the mean value of the observable depending on one
random variable in the form of a correlation function of observables
depending on three random variables.

Analogous representations can be developed for highest moments of
the observable of one random variable.

\section{Examples of $\textit{N}$~=~4 and $\textit{N}$~=~8}
\pst
We recall that in our approach the integer $s$ is {\it the random
variable}, the numbers $p_s$ (the probability distributions) are the
{\it states}, and the function $F(s)$ is {\it the observable}, which
has a value equal to the number $F(s)$. For $s=1,2,\ldots,N$, we
have $N$ values of random variable. One can use any other notation
for the states and random variables, using an invertible map of the
integers $1,2,\ldots,N$ onto another set of numbers
$m_1,m_2,\ldots,m_N$.

\subsection{Case of $\textit{N}$~=~4}
\pst
We study the suggested construction on the example of $N=4$. As an
example, we consider these numbers as numbers associated with a
casino roulette (or geographic compass).

This means that we have four different positions of the casino
roulette $s=1,2,3,4$ (or four directions of the compass arrow).

We use the map $1\leftrightarrow 1,~1;~~2\leftrightarrow
1,~2;~~3\leftrightarrow 2,~1;~~4\leftrightarrow 2,~2$ to label the
four roulette positions by four pairs of numbers $p_{jk}$
$(j,k=1,2)$, i.e., $p_1\equiv p_{11},~p_2\equiv p_{12},~p_3\equiv
p_{21}$, and $~p_4\equiv p_{22}$.

Now we introduce the observable $F(s)$, which is a function of a
random variable $s$ equal to the number $F(s)$ at each value of the
variable. In this way, we have four numbers $~F(s=1)=F(1)$,
$~F(s=2)=F(2)$, $~F(s=3)=F(3)$, and $~F(s=4)=F(4)$.

The mean value of the observable reads
\begin{equation}\label{15}
\langle F\rangle=p_{1}F(1)+p_{2}F(2)+p_{3}F(3)+p_{4}F(4).
\end{equation}
The mean value $\langle F\rangle$ is a functional that depends on
two functions $p_s$ and $F(s)$, i.e., the state and observable.

Using the mapping procedure developed, we can rewrite Eq.~(\ref{15})
as follows:
\begin{equation}\label{16}
\langle F\rangle=p_{11}F(1,1)+p_{12}F(1,2)+p_{21}F(2,1)+p_{22}F(2,2)
\end{equation}
or
\begin{equation}\label{17}
\langle F\rangle=\sum_{j=1}^2\sum_{k=1}^2p_{jk}F(j,k).
\end{equation}
Now we choose the function $F(j,k)$ in the form
\begin{equation}\label{18}
F(1,1)=\varphi(1)\chi(1),\quad F(1,2)=\varphi(1)\chi(2),\quad
F(2,1)=\varphi(2)\chi(1),\quad F(2,2)=\varphi(2)\chi(2).
\end{equation}
One can introduce two other functions $\widetilde\varphi(j,k)$ and
$\widetilde\chi(j,k)$, which provide the same result of
multiplication
\begin{equation}\label{19}
F(1,1)=\widetilde\varphi(1,1)\widetilde\chi(1,1),\quad
F(1,2)=\widetilde\varphi(1,2)\widetilde\chi(1,2),\quad
F(2,1)=\widetilde\varphi(2,1)\widetilde\chi(2,1),\quad
F(2,2)=\widetilde\varphi(2,2)\widetilde\chi(2,2).
\end{equation}
In fact, one should obtain these equalities if
\begin{eqnarray}
\widetilde\varphi(1,1)=\varphi(1),\quad \widetilde\varphi(1,2)=\varphi(1),
\quad \widetilde\varphi(2,1)=\varphi(2),\quad
\widetilde\varphi(2,2)=\varphi(2),\nonumber\\[-2mm]
\label{20}\\[-2mm]
\widetilde\chi(1,1)=\chi(1),\quad \widetilde\chi(1,2)=\chi(1),
\quad \widetilde\chi(2,1)=\chi(2),\quad
\widetilde\chi(2,2)=\chi(2).\nonumber
\end{eqnarray}
We can interpret the functions
$\widetilde\varphi(j,k)\equiv\widetilde\varphi\big(s(j,k)\big)$ and
$\widetilde\chi(j,k)\equiv\widetilde\chi\big(s(j,k)\big)$ as two
specific observables or two different kinds of a function of one
random variable. The results obtained can be summarized as the
equality
\begin{equation}\label{20A}
\langle F\rangle=\langle\widetilde\varphi\widetilde\chi\rangle;
\end{equation}
this means that $\langle F\rangle$, being the classical observable
mean, can be interpreted as the correlation function of two
classical observables $\widetilde\varphi$ and $\widetilde\chi$. We
call the correlations of these two observables $\widetilde\varphi$
and $\widetilde\chi$ depending on one random variable $s$ {\it the
hidden correlations}.

Analogously, for $N=n_1n_2\cdots n_\ell$ one can obtain the equality
\begin{equation}\label{20B}
\langle F\rangle=\langle\widetilde\varphi_1\widetilde\varphi_2
\cdots\widetilde\varphi_\ell\rangle,
\end{equation}
where the same $\langle F\rangle$ can be considered as the
correlation function of $\ell$ observables
$\widetilde\varphi_1,\widetilde\varphi_2,
\ldots,\widetilde\varphi_\ell$ (hidden correlations).

\subsection{Case of $\textit{N}$~=~8}
\pst Now we consider the case of $N=8$, where we also have
nonnegative numbers $p_1,p_2,\ldots p_8$, with $\sum_{s=1}^8p_s=1$.
We use the map $s\leftrightarrow s(j,k,\ell)$, i.e.,
$$ p_1={\cal P}_{111},~~p_2={\cal P}_{112},~~p_3={\cal P}_{121},~~p_4={\cal P}_{122},~~
p_5={\cal P}_{211},~~p_6={\cal P}_{212},~~p_7={\cal
P}_{221},~~p_8={\cal P}_{222}.$$ The nonnegative numbers ${\cal
P}_{jk\ell}$ satisfy the condition $\sum_{j,k,\ell=1}^2{\cal
P}_{jk\ell} =1$. They can be interpreted as a joint probability
distribution of three random variables $j$, $k$, and $\ell$.

We turn to the observable $F(s)$, $s=1,2,\ldots,8$. In terms of the
probability distribution $p_s$, the mean value $\langle F\rangle$
reads $\langle F\rangle=\sum_{s=1}^8F(s)p_s$. In view of the
notation $F\big(s(j,k,\ell)\big)\equiv F(j,k,\ell)$, we arrive at
$$\langle F\rangle=\sum_{j,k,\ell}{\cal P}_{jk\ell}F(j,k,\ell).$$ If
the observable $F(j,k,\ell)$ is taken in the form
$$F(j,k,\ell)=\varphi(j)\chi(k)u(\ell),$$
we obtain $$\langle
F\rangle=\sum_{j,k,\ell}\varphi(j)\chi(k)u(\ell){\cal
P}_{jk\ell}=\langle \varphi(j)\chi(k)u(\ell)\rangle.$$ Thus, we
obtain the result that the mean of a specific observable $F(s)$
appears in the form of the correlation function of three observables
$$\widetilde\varphi(j,k,\ell)=\varphi(j),\quad
\widetilde\chi(j,k,\ell)=\chi(k),\quad \widetilde
u(j,k,\ell)=u(\ell),\qquad\mbox{i.e.},\qquad\langle
F\rangle=\langle\widetilde\varphi\widetilde\chi\widetilde
u\rangle.$$ The functions $\widetilde\varphi$, $\widetilde\chi$, and
$\widetilde u$ can be interpreted as observables depending on one
random variable $s$. Thus, the mean value of the observable $F(s)$
can be written as the correlation function of three observables
$\widetilde\varphi(s)$, $\widetilde\chi(s)$, and $\widetilde u(s)$,
i.e., $$\sum_{s=1}^Np_sF(s) =\sum_{s=1}^Np_s\widetilde\varphi(s)
\widetilde\chi(s)\widetilde u(s).$$

\section{Quantum Qudit States and Observables}
\pst
In this section, we construct quantum states and observables,
extending the approach discussed in the previous sections for
classical systems.

Given $N$$\times$$N$ matrix $\rho_{ss'}$ $(s,s'=1,2,\ldots,N)$. If
$\rho=\rho^\dagger$, $\mbox{Tr}\rho=1$, and $\rho\geq 0$, this
matrix can be interpreted as the density matrix of qudit state with
$j=(N-1)/2$.

At $N=nm$, the matrix $\rho_{ss'}$ $(s,s'=1,2,\ldots,N)$ can be
interpreted as the density matrix of two qudits with $j_1=(n-1)/2$
and $j_2=(m-1)/2$, as well as at $N=n_1n_2n_3$, it can also be
interpreted as the density matrix of three qudits with
$j_1=(n_1-1)/2$, $j_2=(n_2-1)/2$, and $j_3=(n_3-1)/2$. An analogous
interpretation can be provided for $N=\prod_{k=1}^\ell n_k$, and the
matrix $\rho_{ss'}$ $(s,s'=1,2,\ldots,N)$ can be considered as the
density matrix of $\ell$ qudits with $j_k=(n_k-1)/2$.

To provide such an interpretation, we use the map of matrix indices
$s\leftrightarrow j,k,~s'\leftrightarrow j',k'$, i.e., $s=s(j,k)$
and $s'=s'(j',k')$, while considering two qudits, and
$s=s(j,k,\ell)$ and $s'=s'(j',k',\ell')$, while considering three
qudits, etc. We used this tool in \cite{Entropy}. In this paper, we
study the possibility to extend this interpretation also for
matrices of observables $F_{ss'}$ corresponding to the operators
$\hat F$ acting in the Hilbert space ${\cal H}$.

We can write the matrices of observables either in the form
\begin{eqnarray}
F_{ss'}=F_{s(j,k)\,s'(j',k')}\equiv F_{jk,\,j'k'},\label{21}
\end{eqnarray}
or in the form
\begin{eqnarray}
F_{ss'}=F_{s(j,k,\ell)\,s'(j',k',\ell')}\equiv
F_{jk\ell,\,j'k'\ell'},\label{22}
\end{eqnarray}
where indices $j,k$ and $j,k,\ell$ take the same values as in the
density matrix
$$
\rho_{ss'}=\rho_{s(j,k)\,s'(j',\,k')}\equiv \rho_{jk,\,j'k'},\qquad
\rho_{ss'}=\rho_{s(j,k,\ell)\,s'(j',\,k',\,\ell')}\equiv \rho_{jkl,\,j'k'\ell'}.
$$
Thus, both quantum states and quantum observables described by a
density operator $\hat\rho$ and an observable operator $\hat F$
acting in the $N$$\times$$N$-dimensional Hilbert space $\widetilde
H$ can be associated with the matrices $\rho_{ss'}$ and $F_{ss'}$
given in the basis $\mid s\rangle$, i.e., $\rho_{ss'}=\langle
s\mid\hat\rho\mid s'\rangle$ or $F_{ss'}=\langle s\mid\hat F\mid
s'\rangle$.

On the other hand, one can use the basis $\mid s\rangle=\mid
s(j,k)\rangle=\mid j\rangle\mid k\rangle$, considering the Hilbert
space $\widetilde H$ as the tensor product of two Hilbert spaces
$\widetilde H=\widetilde H_1\otimes\widetilde H_2$. In this basis,
the matrix of the same density operator $\hat\rho$ reads
$$\rho_{jk,j'k'}=\langle s(j,k)\mid\hat\rho\mid s'(j',k')\rangle;$$
this is the same numerical $N$$\times$$N$ matrix $\rho_{ss'}$ but
with the matrix elements labeled by indices $jk,\,j'k'$.

Analogously, for the observable $\hat F$ we can write the matrix
$\langle s\mid\hat F\mid s'\rangle$ in the form $$\langle
s(j,k)\mid\hat F\mid s'(j',k')\rangle\equiv F_{jk,\,j'k'}.$$ Thus,
we obtain the same $N$$\times$$N$ numerical matrix with matrix
elements $F_{ss'}$ $(s,s'=1,2,\ldots,N)$ but the matrix elements are
labeled by the indices $jk$ and $j'k'$ $(j,j'=1,2,\ldots,n;~
k,k'=1,2,\ldots,m)$. The map introduced provides a chance to write
the mean value of the observable
$F_{ss'}=F_{s(j,k)\,s'(j',k')}\equiv F_{jk,\,j'k'}$ as
\begin{eqnarray}\label{23}
\langle\hat F\rangle=\mbox{Tr}\,\hat
F\hat\rho&=&\sum_{s=1}^N\sum_{s'=1}^NF_{ss'}\rho_{s's}=\sum_{j=1}^n\sum_{k=1}^m
\sum_{j'=1}^n\sum_{k'=1}^mF_{s(j,k)\,s'(j',k')}\,\rho_{s'(j',k')\,s(j,k)} \nonumber\\
&&= \sum_{j=1}^n\sum_{k=1}^m
\sum_{j'=1}^n\sum_{k'=1}^mF_{jk,j'k'}\,\rho_{j'k',jk}.
\end{eqnarray}

If one takes the observable $\hat F$ in the form
\begin{equation}\label{24}
\hat F=\hat F_1\otimes\hat F_2,
\end{equation}
where $\hat F_1$ is the operator of the observable acting in the
Hilbert space $\widetilde{\cal H}_1$ and $\hat F_2$ is the operator
of the observable acting in the Hilbert space $\widetilde{\cal
H}_2$, Eq.~(\ref{23}) reads
\begin{equation}\label{25}
\langle\hat F\rangle=\sum_{j=1}^n\sum_{k=1}^m
\sum_{j'=1}^n\sum_{k'=1}^m(F_1)_{jj'}(F_2)_{kk'}\,\rho_{j'k',\,jk}.
\end{equation}

In the case where $\hat F=\hat F_1\otimes\hat F_2$, we introduce two
commuting observables
\begin{equation}\label{26}
\widetilde{\hat F}_1=\big(\hat F_1\otimes\hat 1_m\big),\qquad
\widetilde{\hat F}_2 \big(\hat 1_n\otimes\hat F_2\big).
\end{equation}
For these two observables, the mean value of the observable $\hat F$
takes the form of the correlation function of the observables
$\widetilde{\hat F}_1$ and $\widetilde{\hat F}_2$, i.e.,
\begin{equation}\label{27}
\langle\hat F\rangle=\langle\widetilde{\hat F}_1\widetilde{\hat F}_2\rangle.
\end{equation}
As a result, we obtained a quantum analog of the classical
probability relation (\ref{20A}).

For $N=n_1n_2\cdots n_\ell$, we have
\begin{equation}\label{28}
\langle\hat F\rangle=\langle\widetilde{\hat F}_1\widetilde{\hat
F}_2\cdots\widetilde{\hat F}_\ell \rangle;
\end{equation}
this relation is a generalization of Eq.~(\ref{27}). We showed that
the same mean value $\langle\hat F\rangle$ can be considered as the
correlation function of $\ell$ commuting observables
$\widetilde{\hat F}_p$ $(p=1,2,\ldots,\ell)$.

\section{Entropic and Information Inequalities}
\pst
In this section, we consider the classical system with one random
variable. We recall that there exist inequalities for entropies of
joint probability distributions ${\cal P}(j,k)$ of two random
variables $j$ and $k$ of the form
$$-\sum_{jk}{\cal P}(j,k)\ln{\cal P}(j,k)\leq -\sum_j\Big\{\Big[\sum_{k}{\cal P}(j,k)\Big]\ln
\Big[\sum_{k}{\cal P}(j,k)\Big]\Big\}-\sum_{k}\Big\{\Big[\sum_{j}{\cal
P}(j,k)\Big]\ln \Big[\sum_{j}{\cal P}(j,k)\Big]\Big\}.$$ This inequality (the
subadditivity condition) can be interpreted as the subadditivity condition for
the probability distribution of one random variable
$$-\sum_sp_s\ln p_s\leq-\sum_j\Big\{\Big[\sum_{k}p_{s(j,k)}\Big]\ln
\Big[\sum_{k}p_{s(j,k)}\Big]\Big\}-\sum_{k}\Big\{\Big[\sum_{j}p_{s(j,k)}\Big]\ln
\Big[\sum_{j}p_{s(j,k)}\Big]\Big\}.$$

The other example (for $N=4$) is the possibility to consider the
observable $F(j,k)$ for two random variables $j$ and $k$, for
example, $\,F(1,1)=1$, $\,F(1,2)=-1$, $\,F(2,1)=-1$, and
$\,F(2,2)=1$ as an observable for one random variable $F(s)$ such as
$\,F(1)=1$, $\,F(2)=-1$, $\,F(3)=-1$, and $\,F(4)=1$. In this case,
$$\langle F\rangle=p_1F(1)+ p_2F(2)+p_3F(3)+p_4F(4)=p_1-
p_2-p_3+p_4.$$ Also $$\langle F\rangle={\cal P}(1,1)F(1,1)+ {\cal
P}(1,2)F(1,2)+{\cal P}(2,1)F(2,1)+{\cal P}(2,2)F(2,2)={\cal P}(1,1)-
{\cal P}(1,2)-{\cal P}(2,1)+{\cal P}(2,2).$$ On the other hand, for
two observables $$\widetilde F_1(1,1)=1,\quad\widetilde
F_1(1,2)=1,\quad\widetilde F_1(2,1)=-1,\quad\widetilde F_1(2,2)=-1$$
and $$\widetilde F_2(1,1)=1,\quad\widetilde
F_2(1,2)=-1,\quad\widetilde F_2(2,1)=1,\quad\widetilde
F_2(2,2)=-1,$$ one has the correlation function of the form
$$\langle\widetilde F_1\widetilde F_2\rangle=\sum_{j,k}{\cal P}(j,k)\widetilde F_1(j,k)\widetilde
F_2(j,k),$$ and this correlation function is equal to $\langle
F\rangle$. In fact,
\begin{equation}\label{37}
\langle\widetilde F_1\widetilde F_2\rangle={\cal
P}(1,1)- {\cal P}(1,2)-{\cal P}(2,1)+{\cal
P}(2,2)=p_1-p_2-p_3+p_4=\sum_{s=1}^4F(s)p_s.
\end{equation}

In view of the invertibility of the applied map of the indices, we
can introduce two observables $F'_1(s)$ and $F'_2(s)$, i.e.,
$$F'_1(1)=1,\quad F'_1(2)=1,\quad F'_1(3)=-1,\quad F'_1(4)=1$$ and
$$F'_2(1)=1,\quad F'_2(2)=-1,\quad F'_2(3)=1,\quad F'_2(4)=-1.$$
Then one has the equality $\langle F\rangle=\langle F'_1F'_2\rangle$
or
\begin{equation}\label{38}
\sum_{s=1}^4F(s)p_s=\sum_{1=1}^4F'_1(s)F'_2(s)p_s.
\end{equation}
Thus, we showed that the mean value of observable $F(s)$ can be
interpreted as the correlation function of observables $\widetilde
F_1(j,k)$ and $\widetilde F_2(j,k)$. The subadditivity condition for
functions $p_s$ of one random variable reflects the correlations of
two artificial random variables $(j,k)$ that are connected with two
observables $\widetilde F_1$ and $\widetilde F_2$. Another
interpretation of equality~(\ref{38}) reflects the fact that there
exist hidden correlations of observables $F'_1(s)$ ad $F'_2(s)$ for
the case of a single random variable $s$.

\section{Conclusions}
\pst To conclude, we point out the main results of this study.

We showed that for a single qudit with $j=(N-1)/2$ it is possible to
find commuting observables (Hermitian matrices), e.g., two
observables $\hat\varphi$ and $\hat\chi$, such that the product of
the observables provides the Hermitian matrix
$\hat\varphi\hat\chi=\hat A$. Then the mean value of the observable
$\hat A$ can be interpreted as the correlation function of two
observables $\langle\hat
A\rangle=\langle\hat\varphi\hat\chi\rangle=\mbox{Tr}\,\big(\hat
A\hat\rho\big)$.

In the case $N=n_1n_2\cdots n_k$, one can find commuting observables
$\hat\varphi_1,\hat\varphi_2,\ldots,\hat\varphi_k$, such that the
means of the observable $\hat
A=\hat\varphi_1\hat\varphi_2\cdots\hat\varphi_k$ can be treated as
the correlation function $\langle\hat
A\rangle=\langle\hat\varphi_1\hat\varphi_2\cdots\hat\varphi_k\rangle=
\mbox{Tr}\,\big(\hat
A\hat\rho\big)$.

In the case of a multiqudit system with the same numerical density
matrix $\rho(1,2,\ldots,k)$, the observables
$\hat\varphi_1,\hat\varphi_2,\ldots,\hat\varphi_k$ have the physical
meaning of the observables associated with each qudit in the
composite system.

Such an observation means that the quantum correlations known for
observables associated with the subsystems are also available in
single qudit systems like the quantum roulette and the quantum
compass. We call these correlations the hidden correlations and they
can be used in quantum technology applications analogously to the
correlations in composite systems (like, e.g., the entanglement).

\section*{Acknowledgements}
\pst
The authors are grateful to the Organizers of the 22nd Central
European Workshop on Quantum Optics (6--10 July 2015, Warsaw,
Poland) and especially to Prof. K. Banaszek for invitation. M.A.M.
acknowledges the Aleksander Jab{\l}o\'nski Foundation for a visiting
professor fellowship.

\end{document}